\documentclass[preprint]{elsarticle}
\usepackage{lineno,hyperref,pdflscape}
\modulolinenumbers[5]

\usepackage{multicol}
\usepackage{color}
\usepackage{xspace}
\usepackage{enumerate}
\usepackage{amsmath} 
\usepackage{ulem} 
\usepackage{placeins}
\usepackage{threeparttable}

\usepackage{tabularx}
\usepackage{mathtools} % for \mathclap and \substack macros
\usepackage{amsfonts}
\newcolumntype{M}{>{\centering\arraybackslash}m{1.85cm}}
\usepackage[export]{adjustbox}
\usepackage{float}
\usepackage{braket}
\usepackage{lipsum}
\usepackage{longtable}
\usepackage{lscape}
\graphicspath{{figure/}}
\newcommand\T{\rule{0pt}{3ex}}       % Top strut
\newcommand\B{\rule[-1.5ex]{0pt}{0pt}} % Bottom strut

\usepackage{booktabs}
\usepackage{blindtext}
\makeatletter
\newcommand{\colorcaption}[2][]{%
	\begingroup%
	\renewcommand{\@caption@fignum@sep}{ (Color online). }%
	\caption[#1]{#2}%
	\endgroup%
}

\newcommand{\lambdabar}{{\mkern0.75mu\mathchar '26\mkern -9.75mu\lambda}}
\makeatother
\usepackage{amssymb}
\usepackage{graphicx}
\usepackage{rotating}
\usepackage{tikz}
\begin{document}
	
%%%%%%%%%%%%%%%%%%%%%%%%%%%%%%%%%%%%%%%%%%%%%%%%%%%%%%%%%%%%%%%%%
\begin{frontmatter}
\title{Forbidden beta decay properties of $^{135,137}$Te using shell-model}
\author{Shweta Sharma}
\author{Praveen C. Srivastava\footnote{Corresponding author: praveen.srivastava@ph.iitr.ac.in}}
\author{Anil Kumar}
\address{Department of Physics, Indian Institute of Technology Roorkee, Roorkee 247667, India}

\date{\hfill \today}
%%%%%%%%%%%%%%%%%%%%%%%%%%%%%%%%%%%%%%%%%
\begin{abstract}

In this work, the large-scale shell-model calculations for $\beta$-decay properties have been done. The $\beta$-delayed $\gamma$-ray spectroscopy has been performed recently at ILL, Grenoble [M. Si \textit{et al.}, Phys. Rev. C {\bf 106}, 014302 (2022)] to study $\beta$-decay properties corresponding to $^{137}$Te (($7/2^-))$ $\rightarrow$  $^{137}$I ($J_f$) transitions. 
We have done systematic shell-model study for nuclear structure properties and compared the obtained results with the experimental data. Finally, the $\beta$-decay properties such as the $\log ft$ values and average shape factors have been reported. This is the first theoretical calculation for the $\log ft$ values corresponding to these new experimental data. In addition, we have also reported calculated $\log ft$ results for $^{135}$Te (($7/2^-))$ $\rightarrow$  $^{135}$I ($J_f$) transitions.
\end{abstract}
\begin{keyword}
Shell-model, forbidden beta decay. 
\end{keyword}
\end{frontmatter}
	
	%\maketitle
	
	%%%%%%%%%%%%%%%%%%%%%%%%%%%%%%%%%%%%%%%%%%%%%%%%%%%%%%%%%%%%%%%%%%%%%%%%%

%%%%%%%%%%%%%%%%%%%%%%%%%%%%%%%%%%%%%%%%%%%%%%%%%%%%%%%%%%%%%%%%%%%%%%%%%

\section{Introduction}
The nuclear shell-model is the most desirable and trustworthy in calculating nuclear structure properties. Nuclei with the magic number of the proton or neutron number play an essential role in determining the elemental abundance in the universe.   
Exploring the nuclear structure properties in the vicinity of doubly magic nuclei is an important task. Further, if we move towards heavier neutron-rich nuclei,  the evaluation of the nuclear structure and beta decay properties of such nuclei is a difficult task both experimentally and theoretically. However, despite the complexities associated, these neutron-rich nuclei attract considerable attention because of astrophysical aspects, i.e., $r$-process \cite{rprocess} path modeling. One such example of heavier doubly magic neutron-rich nuclei is $^{132}$Sn with $Z=50$ and $N=82$. Moreover, it has been found that $^{132}$Sn shows robust shell closure similar to $^{208}$Pb \cite{coraggio,blom}. Recently, the beta-decay properties of nuclei in the vicinity of doubly magic nucleus $^{208}$Pb are calculated with the help of large-scale shell-model calculations by our group \cite{sharma,anil21npa} and good results are obtained, which shows the reliability of shell-model. Further, the nuclear shell-model has been used in the evaluation of nuclear structure properties in $^{132}$Sn region \cite{saha}. Thus, studying $^{135,137}$Te using a systematic shell-model study will aid the valuable information in the presently available experimental data.

These neutron-rich $r$-process nuclei show beta-decay, which is an electroweak process changing the nuclear charge number by one unit, making this decay the major decay process in this region. These weak processes can be characterized in various types, i.e., allowed $(l=0)$ and forbidden $(l>0)$ beta decay based on the angular momentum of emitted leptons. Further, forbidden beta decay can be characterized by two types: Forbidden unique (FU) beta decay and forbidden non-unique (FNU) beta decay. In FU beta decay, the total change in angular momentum, i.e., $\Delta J$ is equal to $l+1$, whereas, in the case of FNU beta decay, $\Delta J$ is equal to $l-1,l$ where $l$ denotes the degree of forbiddenness.
Competition between allowed and forbidden beta decay in the higher mass region makes the evaluation of beta decay properties interesting \cite{car,pcs}. Further, the values of weak coupling constants are quenched in large-scale shell-model calculations because of many nuclear medium effects \cite{jouni17}, and this quenching increases as we move towards higher mass region. Moreover, the value of rank-0 nuclear matrix element $\gamma_5$ for first-forbidden beta decay with $\Delta J=0^-$ gets enhanced over its impulse-approximation value with the help of mesonic enhancement factor  \cite{Towner} ($\epsilon_{MEC}$). The effect of mesonic enhancement factor on half-lives and spectral shapes of first forbidden beta decay is shown in Ref. \cite{joel} by considering three sets of enhancement factors, i.e., $\epsilon_{MEC}=1.4,1.7,2.0$ for $A\sim 135$ region.

Recently, Si \textit{et al.} \cite{si} studied $\gamma$-ray spectroscopy of $^{137}$Te at ILL, Grenoble using LOHENGRIN recoil separator and populated $^{137}$Te to study ground and several excited states of $^{137}$I. Further, they have measured $\beta$-decay properties like $\log ft$ values and branching fractions corresponding to these transitions. The total half-life evaluated was 2.46(5) s along with $\beta$-delayed neutron emission probability of 2.63(85)$\%$. 
Further, in this work they have also compared experimental energy spectra with shell-model corresponding to N3LOP interaction. 

These results motivated us to report  $\beta$-decay properties like $\log ft$ values and average shape factors for these transitions using systematic shell-model calculations for the first time. We have also calculated energy spectra and main wave function configurations using jj56pnb effective interaction. The electromagnetic properties of these nuclei have been evaluated for a better understanding of the nuclear structure properties of these nuclei.  Along with $^{137}$Te, the beta decay properties of $^{135}$Te have also been studied in the present paper. 

We have organized our work  as follows: Formalism about shell-model and beta decay have been given in section \ref{formalism}, and results are discussed in section \ref{secIII}, which includes nuclear structure properties like energy spectra, wave function, and electromagnetic properties. Beta decay properties such as $\log ft$ values and average shape factors have been discussed in this section. Finally, the paper is concluded in section \ref{secIV}.

\section{Formalism} \label{formalism}

\subsection{{\bf Shell-model Hamiltonian}}

In the present paper, the large-scale shell-model calculations have been used to determine nuclear structure and beta decay properties. The nuclear shell-model Hamiltonian \cite{jouni} containing one and two body components can be written as

\begin{equation}
H = T + V = \sum_{\alpha}{\epsilon}_{\alpha} c^{\dagger}_{\alpha} c_{\alpha} + \frac{1}{4} \sum_{\alpha\beta \gamma \delta}v_{\alpha \beta \gamma \delta} c^{\dagger}_{\alpha} c^{\dagger}_{\beta} c_{\delta} c_{\gamma},
\end{equation}
where $T$ is the kinetic energy term, and $V$ is the potential energy term. The term $\alpha = \{n,l,j,t\}$ stands for single-particle state where $n,l,j,t$ are principal, orbital, total angular momentum and isospin quantum numbers, respectively and $\epsilon_{\alpha}$ is the corresponding single particle energy. $v_{\alpha \beta \gamma \delta} = \langle\alpha \beta | V | \gamma \delta\rangle $ stands for the antisymmetrized two-body matrix elements. $c^{\dagger}_{\alpha}$ and $c_{\alpha}$ are creation and annihilation operators for nucleons, respectively. 

\subsection{ {\bf $\beta$-decay theory}}

Beta decay formalism using impulse approximation is explained here in brief. The detailed formalism can be found in Refs. \cite{beh82,hfs}. In beta decay, the total half-life of the nucleus can be calculated with the help of partial half-lives of all the transitions in that nucleus. Thus, the partial half-life is given by
\begin{eqnarray}\label{hf1}
t_{1/2}=\frac{\kappa}{\tilde{C}},
\end{eqnarray}
where $\kappa$ \cite{patri} is the universal constant value and $\tilde{C}$ is the dimensionless integrated shape function. The value of $\kappa$ is given by

\begin{eqnarray}
\kappa=\frac{2\pi^3\hbar^7\text{ln(2)}}{m_e^5c^4(G_\text{F}\text{Cos}\theta_\text{C})^2}=6289~\mathrm{s},
\end{eqnarray} 
where $\theta_\text{C}$ stands for Cabibbo angle and $m_e$ is the electron mass. The term $G_F$ denotes the Fermi constant, i.e., effective coupling constant. Further, $\tilde{C}$ can be written as

\begin{eqnarray} \label{tc}
\tilde{C}=\int_1^{w_0}C(w_e)pw_e(w_0-w_e)^2F_0(Z,w_e)dw_e.
\end{eqnarray}

The terms $w_0=W_0/m_ec^2$, $w_e=W_e/m_ec^2$, and $p=p_ec/m_ec^2=\sqrt{(w_e^2-1)}$ are dimensionless kinematic quantities. Here, $W_e$ and $p_e$ are emitted lepton energy and momentum, respectively. $W_0$ is the end-point energy. The term $C(w_e)$ is the shape factor that depends on electron energy, and $F_0(Z,w_e)$ is the Fermi function introduced because of the Coulombic force between the beta particle and nucleus. Further, in the case of allowed transition, $C(w_e)=B(GT)+B(F)$ where B(GT) denotes the Gamow-Teller reduced transition probability and B(F) denotes the Fermi reduced transition probability. 
We can define B(F) and B(GT) as 
\begin{equation}
B(F) = \frac{g_V^2}{2J_{i}+1}|M_F|^2,~~  B(GT) = \frac{g_A^2}{2J_{i}+1}|M_{GT}|^2,
\end{equation}
where $J_{i}$ is the angular momentum of the initial nuclear state, and $g_{V}$ and $g_{A}$ are the vector and axial-vector coupling constants, respectively. $M_{F}$ and $M_{GT}$ are the matrix elements corresponding to Fermi and Gamow-Teller transitions, respectively \cite{jouni}.

The phase space factor is given by

\begin{equation}
f_{0}= \int_1^{w_0}pw_e(w_0-w_e)^2F_0(Z,w_e)dw_e.
\end{equation}

In the case of forbidden beta decay, the shape factor is given by
\begin{align} \label{eq11}
\begin{split}
C(w_e)  = \sum_{k_e,k_\nu,K}\lambda_{k_e} \Big[M_K(k_e,k_\nu)^2+m_K(k_e,k_\nu)^2 \\
-\frac{2\gamma_{k_e}}{k_ew_e}M_K(k_e,k_\nu)m_K(k_e,k_\nu)\Big],
\end{split}
\end{align}
where $K$ denotes the forbiddenness order and $k_e$ and $k_\nu$ are the positive integers emerging from the partial wave expansion of the leptonic wave function. The terms $\gamma_{k_e}$ and $\lambda_{k_e}$ are the auxiliary quantities. Further the quantities $M_K(k_e,k_\nu)$ and $m_K(k_e,k_\nu)$ are the complicated combinations of nuclear form factors and leptonic phase space factors. The detailed explanation of these expressions can be found in \cite{mika2017}. These form factors are written in the form of nuclear matrix elements \cite{Anil,anil21} such that

\begin{align}
	\begin{split}
		^{V/A}\mathcal{M}_{KLS}^{(N)}(pn)(k_e,m,n,\rho)& \\ =\frac{\sqrt{4\pi}}{\widehat{J}_i}
		\sum_{pn} \, ^{V/A}m_{KLS}^{(N)}(pn)(&k_e,m,n,\rho)(\Psi_f\parallel [c_p^{\dagger}\tilde{c}_n]_K\parallel \Psi_i),
	\end{split}
\end{align}
where $^{V/A}m_{KLS}^{(N)}(pn)(k_e,m,n,\rho)$ denotes the single particle matrix elements and the quantity $(\Psi_f\parallel [c_p^{\dagger}\tilde{c}_n]_K\parallel \Psi_i)$ denotes the one body transition density (OBTD) which is obtained from the NUSHELLX \cite{nushellx} and KSHELL \cite{shim19}. Once we find $f_0$ and $t_{1/2}$, we can denote it in the form of $\log ft$ values because half-life values can be very large.

\begin{equation}\label{eq9}
	\mbox{log} ft \equiv \mbox{log}(f_{0}t_{1/2}).
\end{equation}

In the case of forbidden unique transitions, this $\log ft$ value is given as
\begin{equation}\label{eq10}
\mbox{log} ft \equiv \mbox{log}(f_{Ku}t_{1/2}),
\end{equation}
where $f_{Ku}$ is the phase-space factor for $K^{\text{th}}$ forbidden unique transition \cite{jouni17} which is given by
\begin{equation}
f_{Ku}= \Big(\frac{3}{4}\Big)^K\frac{(2K)!!}{(2K+1)!!}\int_1^{w_0}C_{Ku}(w_e)p_e w_e(w_0-w_e)^2F_0(Z_f,w_e)dw_e,
\end{equation}
where $C_{Ku}(w_e)$ is the shape function which is given as
\begin{equation}
C_{Ku}(w_e)=\sum_{k_e+k_{\nu}=K+2}\frac{\lambda_{k_e}p_e^{2(k_e-1)}(w_0-w_e)^{2(k_{\nu}-1)}}{(2k_e-1)!(2k_{\nu}-1)!} .
\end{equation}

Here, $\lambda_{k_e}=F_{k_e-1}(Z,w_e)/F_0(Z,w_e)$ where $F_{k_e-1}(Z,w_e)$ is the generalized Fermi function which is given as
\begin{eqnarray}
F_{k_e-1}(Z,w_e) &=4^{k_e-1}(2k_e)(k_e+\gamma_{k_e})[(2k_e-1)!!]^2e^{\pi{y}} \nonumber \\
& \times\left(\frac{2p_eR}{\hbar}\right)^{2(\gamma_{k_e}-k_e)}\left(\frac{|\Gamma(\gamma_{k_e}+iy)|}{\Gamma(1+2\gamma_{k_e})}\right)^2.
\end{eqnarray}

Here, $y=(\alpha{Zw_e}/p_ec)$, where $\alpha=1/137$ is the fine structure constant.

Further, for first forbidden unique transitions, $f_{1u}=12 f_{K=1,u}$.
Additionally, according to Behrens and B$\ddot{\text{u}}$hring \cite{beh82} the average shape factor \cite{enhancement,pri21} for beta-decay can be computed with the help of $ft$ values which is provided by

\begin{equation}
\overline{C(w_{e})} (\text{fm}^{2n}) =\frac{6289 \lambdabar_{\text{Ce}}^{2n}}{ft},
\end{equation} 
where  $\lambdabar_{\text{Ce}}=386.15926796$ fm is the reduced Compton wavelength of electron and $ft$ denotes the reduced half-life and can be evaluated from eqns. \ref{eq9} and \ref{eq10}. Here, $n$ denotes the $n^{\text{th}}$ forbidden beta decay.

\begin{figure*}
	\includegraphics[scale=2.0]{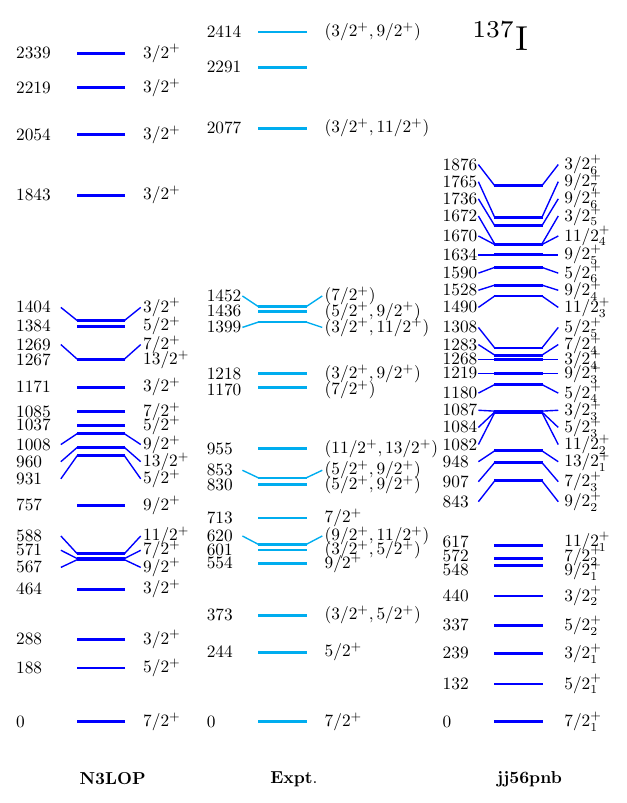}
	%\label{fig:spectra}
	\caption{\label{fig:spectra} Comparison between theoretical and experimental \cite{si}  energy spectra for ground and several excited states of $^{137}$I.  For comparison we have also shown the shell model results corresponding to N3LOP interaction as reported in Ref. \cite{si}.}
\end{figure*}

\section{Results and Discussion} \label{secIII}

\begin{table*}
	 \centering
	\caption{Dominant wave function configurations and probability calculated from large scale shell-model calculations for ground and excited states of $^{137}$I.}\label{table1} 
%	\begin{ruledtabular}
		\begin{tabular}{ccc}
			\hline
			$J_i^{\pi}$ (SM)	& Configuration  & Probability($\%$)     \T\B\\\hline\T\B
			$7/2^+_1$ & $\pi(g^3_{7/2})\nu(f^2_{7/2})$ & 22.84\\ \T\B
			$5/2^+_1$ & $\pi(g^2_{7/2}d_{5/2})\nu(f^2_{7/2})$ & 41.37\\ \T\B
			$3/2^+_1$ & $\pi(g_{7/2}d^2_{5/2})\nu(f^2_{7/2})$ & 9.87\\ \T\B	
			$5/2^+_2$ & $\pi(g^3_{7/2})\nu(f^2_{7/2})$ & 24.62\\ \T\B
			$9/2^+_1$ & $\pi(g^2_{7/2}d_{5/2})\nu(f^2_{7/2})$ & 19.72\\ \T\B
			$3/2^+_2$ & $\pi(g^3_{7/2})\nu(f^2_{7/2})$ & 14.70\\ \T\B
			$5/2^+_3$ & $\pi(g^3_{7/2})\nu(f^2_{7/2})$ & 15.14\\ \T\B
			$9/2^+_2$ & $\pi(g^2_{7/2}d_{5/2})\nu(f^2_{7/2})$ & 20.93\\ \T\B
			$11/2^+_1$ & $\pi(g_{7/2}d^2_{5/2})\nu(f^2_{7/2})$ & 13.89\\ \T\B
			$7/2^+_2$ & $\pi(g^3_{7/2})\nu(f^2_{7/2})$ & 14.05\\ \T\B
			$5/2^+_4$ & $\pi(g^2_{7/2}d_{5/2})\nu(f^2_{7/2})$ & 24.78\\ \T\B
			$9/2^+_3$ & $\pi(g^3_{7/2})\nu(f^2_{7/2})$ & 37.92\\ \T\B
			$5/2^+_5$ & $\pi(g^3_{7/2})\nu(f^2_{7/2})$ & 29.86\\ \T\B
			$9/2^+_4$ & $\pi(g^3_{7/2})\nu(f^2_{7/2})$ & 31.18\\ \T\B
			$11/2^+_2$ & $\pi(g^3_{7/2})\nu(f^2_{7/2})$ & 15.16\\ \T\B
			$13/2^+_1$ & $\pi(g^2_{7/2}d_{5/2})\nu(f^2_{7/2})$ & 22.49\\ \T\B
			$7/2^+_3$ & $\pi(g^2_{7/2}d_{5/2})\nu(f^2_{7/2})$ & 21.79\\ \T\B
			$3/2^+_3$ & $\pi(g^3_{7/2})\nu(f^2_{7/2})$ & 16.35\\ \T\B
			$9/2^+_5$ & $\pi(g^2_{7/2}d_{5/2})\nu(f^2_{7/2})$ & 18.05\\ \T\B
			$3/2^+_4$ & $\pi(g^2_{7/2}d_{5/2})\nu(f^2_{7/2})$ & 15.29\\ \T\B
			$11/2^+_3$ & $\pi(g^3_{7/2})\nu(f^2_{7/2})$ & 22.07\\ \T\B
			$5/2^+_6$ & $\pi(g^2_{7/2}d_{5/2})\nu(f^2_{7/2})$ & 28.60\\ \T\B
			$9/2^+_6$ & $\pi(g^3_{7/2})\nu(f^2_{7/2})$ & 19.01\\ \T\B	
			$7/2^+_4$ & $\pi(g^2_{7/2}d_{5/2})\nu(f^2_{7/2})$ & 27.26\\ \T\B
			$3/2^+_5$ & $\pi(g^3_{7/2})\nu(f^2_{7/2})$ & 43.02\\ \T\B
			$11/2^+_4$ & $\pi(g^3_{7/2})\nu(f^2_{7/2})$ & 43.90\\ \T\B
			$3/2^+_6$ & $\pi(g^3_{7/2})\nu(f^2_{7/2})$ & 17.27\\ \T\B
			$9/2^+_7$ & $\pi(g^2_{7/2}d_{5/2})\nu(f^2_{7/2})$ & 23.54\\
			\hline
		\end{tabular}
%	\end{ruledtabular}
\end{table*}

Large-scale shell-model calculations are used to determine the wave function in order to obtain beta decay results for $^{135,137}$Te nuclei. The $^{132}$Sn ($Z=50,N=82$) nucleus is treated as the core for this purpose. The interaction considered here is jj56pnb \cite{brown} which covers proton number $Z=50-82$ and neutron number $N=82-126$ model space. This interaction is derived from N3LO interaction \cite{n3lo} and the model space include orbitals $0g_{7/2}, 1d_{5/2}, 1d_{3/2}, 2s_{1/2}, 0h_{11/2}$ for protons and $0h_{9/2}, 1f_{7/2}, 1f_{5/2}, 2p_{3/2}, 2p_{1/2}, 0i_{13/2}$ for neutrons.  Further, NUSHELLX and KSHELL are used for the calculation of shell-model observables like wave function and OBTDs. 

\subsection{ {\bf Energy levels  and electromagnetic properties}}

In order to test the aforementioned jj56pnb effective interaction, we first calculated the energy spectra and wave functions corresponding to the spin parity states of these nuclei before computing the beta decay properties for the same. The energy spectra of ground and several excited states of $^{137}$I are shown in Fig. \ref{fig:spectra}.   In this figure, we have also compared our shell-model results corresponding to jj56pnb with N3LOP as reported in Ref. \cite{si}. When jj56pnb and N3LOP interactions are compared, the N3LOP interaction reproduces results that are closer to the experimental values in most of the cases, as compared to the jj56pnb interaction. For instance, the N3LOP interaction predicts the first excited state $5/2^+$ with a difference of 56 keV with experimental
	data, while the difference is 112 keV with jj56pnb interaction. However in some cases, the jj56pnb interaction reproduces better results as compared to the N3LOP ineraction. For instance, the N3LOP interaction predicts the $9/2^+_1$ state with a difference of 19 keV with experimental
	data, while the difference is 6 keV with jj56pnb interaction. Thus, both of the ineractions are suitable for calculation of beta decay properties in this case. However, in the present work, we have used jj56pnb interaction for beta decay properties calculations and energy spectra using jj56pnb interaction is discussed below.
	
	The ground state energy of spin parity state $7/2_1^+$ is well reproduced with the help of this shell-model interaction.  Both experimental and shell-model calculations predict $7/2^-$ and $7/2_1^+$ to be the ground states of $^{137}$Te and $^{137}$I, respectively and similar results are predicted for $^{135}$Te($7/2_{gs}^-$)  and $^{135}$I($7/2_{gs}^+$). Also, these spectra are well reproduced for some of the excited spin parity states of $^{137}$I. For instance, for the spin parity state $9/2_1^+$, the energy given by the shell-model is 548 keV which is near to the experimental value, i.e., 554.2(10) keV. Similarly, in the case of spin parity state $11/2_1^+$, the shell-model calculated energy is 617 keV, whereas the experimental value is 620.5(10) keV. Although, this spin parity state is unconfirmed experimentally. Further, moving upward the energy levels, the experimental energy value for the unconfirmed spin parity state  $13/2_1^+$ is 955.0(12) keV near to shell-model energy value, i.e., 948 keV.  

 %Overall N3LOP result seems more close to the experimental data in comparison to jj56pnb interaction results.}

\begin{table*}
	% \centering
	\caption{ Comparison between experimental \cite{iaea} and theoretically calculated electromagnetic observables corresponding to $^{135,137}$I.}\label{table2} 
%	\begin{ruledtabular}
		\begin{tabular}{ccccccccccc}
			\hline\hline
			Nucleus&\multicolumn{2}{c}{~~~~~$\mu(7/2^+_1)(\mu_N)$} & \multicolumn{2}{c}{~~~~~$Q(7/2^+_1)(eb)$} \T\B \\
			\cline{2-3}
			\cline{4-5}
			&	Expt. & SM&  Expt. & SM      \T\B\\\hline\hline\T\B
			$^{135}$I	&2.940(2) & 2.372&  NA & -0.133 \T\B\\\hline\T\B
			$^{137}$I	&NA & 2.114&  NA & -0.425\\
			\hline\hline
				&\multicolumn{2}{c}{~~~~~$B(E2:5/2^+_1 \rightarrow 7/2^+_1)(e^2fm^4)$} & \multicolumn{2}{c}{~~~~~$ B(M1:5/2^+_1 \rightarrow 7/2^+_1)(\mu_N^2)$} \T\B \\
			\cline{2-3}
			\cline{4-5}
			&	Expt. & SM&  Expt. & SM      \T\B\\\hline\hline\T\B
			$^{135}$I	 & NA & 52.101& NA & 0.00004\T\B\\\hline\T\B
			$^{137}$I	& NA & 1.930& NA & 0.0002\\
			\hline\hline
		\end{tabular}
%	\end{ruledtabular}
\end{table*}

%\subsection{Wave-function configuration}

Table \ref{table1} shows the possible nuclear configurations with probability for ground and several excited states of $^{137}$I. The $7/2_1^+$ and $5/2_1^+$ states are single-particle in nature  with configurations $\pi g_{7/2}^1$ and 
$\pi d_{5/2}^1$, respectively.
Further, it can be inferred that the orbital $d_{5/2}$  contributes to proton configurations along with $g_{7/2}$ orbital in most of the cases. In contrast, for neutrons, only the orbital $f_{7/2}$ contributes in the case of the most probable configurations which means there is no neutron excitation to the higher orbitals.

%\subsection{Electromagnetic observables}

As pointed out in Ref. \cite{si}, the $^{137}$I behaves analogously like $^{135}$I, therefore, it is interesting to calculate nuclear structure properties of both of the transitions and compare them. Thus, in Table \ref{table2}, electromagnetic observables have been reported for both $^{135}$I and $^{137}$I. The magnetic moment for $^{135}$I$(7/2^+_1)$ comes out to be 2.372 $\mu_N$ using shell-model calculations, which is close to the experimental value, i.e., 2.940(2) $\mu_N$. As no experimental data is available for other electromagnetic observables corresponding to these nuclei, thus, theoretical estimates have been given using large-scale shell-model calculations. The theoretical quadrupole moment value for $^{135}$I$(7/2^+_1)$ is -0.133 $eb$ while $B(E2)$ and $B(M1)$ values corresponding to $^{135}$I$(5/2^+_1 \rightarrow 7/2^+_1)$ transition are 52.101 $e^2fm^4$ and 0.00004 $\mu_N^2$, respectively. Further, the magnetic moment value for $^{137}$I$(7/2^+_1)$ using shell-model comes out to be 2.114 $\mu_N$, while the quadrupole moment value is -0.425 $eb$. In addition to this, the $B(E2)$ value for transition $^{137}$I$(5/2^+_1 \rightarrow 7/2^+_1)$ is 1.930 $e^2fm^4$ and $B(M1)$ value for the same transition is 0.0002 $\mu_N^2$.

\subsection{ {\bf The $\log ft$ values}}

Table \ref{table3} compares experimental and theoretical $\log ft$ and average shape factor values corresponding to $^{137}$Te (($7/2^-))$ $\rightarrow$  $^{137}$I ($J_f$) transitions.   The spin parities in double brackets show unconfirmed states. These calculations have been performed using quenching in the axial vector coupling constant, i.e., $g_A^{eff}= 0.66\pm 0.03$ and mesonic enhancement factor, i.e., $\epsilon_{MEC}=2.0$, both taken from Ref \cite{joel}. It can be gathered from table \ref{table3} that the shell-model $\log ft$ values show reasonable agreement with the experimental ones. For instance, for ground state transition, from $^{137}$Te$((7/2^-))$ to $^{137}$I$(7/2_1^+)$ the shell-model $\log ft$ value is 5.712, which is very near to the experimental $\log ft$ value, i.e., 5.8(1). Further, in the case of $^{137}$Te$((7/2^-))$ to $^{137}$I$(5/2_1^+)$ transition at energy 243.6(8) keV, the shell-model $\log ft$ value is 6.045 near to the experimental value, i.e., 6.2(1). Also, in the case of $^{137}$Te$((7/2^-))$ to $^{137}$I$(9/2_1^+)$ transition at energy 554.2(10) keV, the experimental $\log ft$ value, i.e., 6.4(1) is close to the theoretical $\log ft$ value, i.e., 6.976. Moving forward for $^{137}$Te$((7/2^-))$ to $^{137}$I$((7/2_2^+))$ transition at energy 713.5(7) keV, the shell-model $\log ft$ value, i.e., 6.1 and the experimental $\log ft$ values, i.e., 6.0(1) are very close to each other. In similar manner, we can conclude for $^{137}$Te$((7/2^-))$ to $^{137}$I$((7/2_4^+))$ transition at energy 1451.9(7) keV, that shell-model $\log ft$ value, i.e., 6.810 agree pretty well with the experimental value, i.e., 6.6(1). This discussion shows the credibility and validity of shell-model calculations in the case of beta decay properties.

\begin{landscape}
%\begin{sidewaystable}
%\tablewidth=453.6pt
\begin{table*}
	 \centering
	\caption{ Comparison between theoretical and experimental \cite{si} $\log ft$ values for $^{137}$Te$((7/2^-))$ to the ground and different excited states in $^{137}$I transitions. Here, symbol `*' indicates the spin parity states proposed by the shell-model  and `1u' denotes 
	first forbidden unique beta decay with log$ft$ = log$f_{1u}t$. }\label{table3} 
%	\begin{ruledtabular}
		\begin{tabular}{lccc|cccccccc}
			\hline
			\multicolumn{4}{c}{~~~~Expt.}&\multicolumn{4}{c}{~~~~~SM}\T\B \\
			\cline{1-4}
			\cline{5-8}
					
			$J_f^{\pi}$ & Energy& $\log ft$& $ {[\overline{C(w_{e})}]}^{1/2}$& $J_f^{\pi}$& Decay mode&$\log ft$ & $ {[\overline{C(w_{e})}]}^{1/2}$      \T\B\\\hline\T\B
			
			$7/2^+$ & 0.0&5.8(1)&38.553&$7/2^+_1$ & 1st FNU	&5.712  & 42.640 \\ \T\B
			$5/2^+$ & 243.6(8)&6.2(1)&24.325&$5/2^+_1$ & 1st FNU	&6.045  & 29.069 \\ \T\B
			$(3/2^+,5/2^+)$&373.1(7) & 6.6(2)& 15.348&$3/2^+_1$& 1st FU & 11.327  &0.066  \\ \T\B
			$(3/2^+,5/2^+)$	& 373.1(7)&6.6(2)& 15.348 &$5/2^+_2$*&1st FNU& 6.435  & 18.558 \\ \T\B
			$9/2^+$ 	& 554.2(10)&6.4(1) & 19.322& $9/2^+_1$ & 1st FNU	& 6.976 & 9.950 \\ \T\B
			$(3/2^+,5/2^+)$& 600.6(6)&6.8(1) & 12.191  &$3/2^+_2$ & 1st FU& 9.007 & 0.960  \\ \T\B
			$(3/2^+,5/2^+)$ &600.6(6)	& 6.8(1)& 12.191&$5/2^+_3$* & 1st FNU	& 5.980  & 31.343 \\ \T\B
			$(9/2^+,11/2^+)$ &620.5(10)	& 6.8(1)&12.191& $9/2^+_2$*&1st FNU	& 7.688  &  4.384\\ \T\B
			$(9/2^+,11/2^+)$	&620.5(10)& 6.8(1)&12.191&   $11/2^+_1$&1st FU  & 9.261 &  0.717 \\ \T\B
			$(7/2^+)$& 713.5(7)&6.0(1) & 30.624 & $7/2^+_2$ & 1st FNU		& 6.100 & 27.287  \\ \T\B
			$(5/2^+,9/2^+)$ &830.2(6)& 7.0(1) & 9.684&$5/2^+_4$& 1st FNU	& 6.050 & 28.908\\ \T\B
			$(5/2^+,9/2^+)$&830.2(6)& 7.0(1)& 9.684 &  $9/2^+_3$	& 1st FNU& 8.038  & 2.930 \\ \T\B
			$(5/2^+,9/2^+)$ &853.2(7)& 7.0(1)& 9.684&$5/2^+_5$*& 1st FNU	& 6.393  & 19.468 \\ \T\B
			$(5/2^+,9/2^+)$&853.2(7)& 7.0(1) & 9.684&  $9/2^+_4$	& 1st FNU & 8.986 & 0.984 \\ \T\B
			$(11/2^+,13/2^+)$ &955.0(12)& 7.6-9.6$^{1u}$ & 4.853-0.485$^{1u}$ &$11/2^+_2$*& 1st FU	&  8.158-10.134$^{1u}$ & 2.553-0.262$^{1u}$ \\ \T\B
			$(11/2^+,13/2^+)$&955.0(12)& 7.6-9.6$^{1u}$&   723752.812-72375.242$^{1u}$ &$13/2^+_1$	& 3rd FNU & 12.550 & 2424.321\\ \T\B
			$(7/2^+)$	& 1170.1(6)&6.6(1)& 15.348& $7/2^+_3$& 1st FNU	& 7.867  &  3.568 \\ \hline
					\end{tabular}
					%\footnote{xy}
		%	\end{ruledtabular}
	\end{table*}
	%\end{sidewaystable}
\addtocounter{table}{-1}

%\begin{sidewaystable}
\begin{table*}
	\leavevmode
	 \centering
	\caption{ Continued.}\label{table3} 
	%	\begin{ruledtabular}
	\begin{tabular}{lccc|cccccccc}
		\hline
		\multicolumn{4}{c}{~~~~Expt.}&\multicolumn{4}{c}{~~~~~SM}\T\B \\
		\cline{1-4}
		\cline{5-8}
		
		$J_f^{\pi}$ & Energy& $\log ft$& $ {[\overline{C(w_{e})}]}^{1/2}$& $J_f^{\pi}$& Decay mode&$\log ft$ & $ {[\overline{C(w_{e})}]}^{1/2}$      \T\B\\\hline\T\B

	%	$J_f^{\pi}$ (Expt.) & $J_f^{\pi}$ (SM) & & & Expt. & SM & Expt. & SM     \T\B\\\hline\T\B
			$(3/2^+,9/2^+)$	&1218.0(11)& 7.5(1)&5.446&  $3/2^+_3$ & 1st FU	& 9.114   & 0.850   \\ \T\B
			$(3/2^+,9/2^+)$ 	& 1218.0(11)&7.5(1)& 5.446	&  $9/2^+_5$ &1st FNU&6.129 & 26.405 \\ \T\B
			$(3/2^+,11/2^+)$	&1399.0(11)& 7.7-9.6$^{1u}$&4.326-0.485$^{1u}$&  $3/2^+_4$ & 1st FU	& 9.368-11.283$^{1u}$   & 0.634-0.070$^{1u}$  \\ \T\B
			$(3/2^+,11/2^+)$ 	& 1399.0(11)&7.7-9.6$^{1u}$& 4.326-0.485$^{1u}$&  $11/2^+_3$ &1st FU	&9.932-11.846$^{1u}$ & 0.331-0.037$^{1u}$ \\ \T\B
			$(5/2^+,9/2^+)$	&1435.9(9)& 7.1(1)& 8.631 &  $5/2^+_6$ & 1st FNU	& 7.596   & 4.877  \\ \T\B
			$(5/2^+,9/2^+)$ 	& 1435.9(9)&7.1(1)& 8.631 &  $9/2^+_6$* &1st FNU	&6.789 & 12.350 \\ \T\B
			$(7/2^+)$	& 1451.9(7)&6.6(1) & 15.348& $7/2^+_4$& 1st FNU	& 6.810 &  12.059 \\ \T\B
			$(3/2^+,11/2^+)$	&2077.3(14)& 7.7-9.5$^{1u}$&4.326-0.546$^{1u}$ &  $3/2^+_5$ & 1st FU	& 8.818-10.629$^{1u}$  & 1.194-0.148$^{1u}$   \\ \T\B
			$(3/2^+,11/2^+)$	& 2077.3(14)&7.7-9.5$^{1u}$& 4.326-0.546$^{1u}$&  $11/2^+_4$ &1st FU 	& 9.302-11.113$^{1u}$ & 0.684-0.085$^{1u}$ \\ \T\B
			$(3/2^+,9/2^+)$	&2413.8(12)& 7.3(1)	&6.856&  $3/2^+_6$ & 1st FU& 8.670   & 1.416   \\ \T\B
			$(3/2^+,9/2^+)$ 	& 2413.8(12)&7.3(1)& 6.856&  $9/2^+_7$ &1st FNU	&8.784 & 1.242\\ \hline
			
		\end{tabular}
%	\end{ruledtabular}
\end{table*}
%\end{sidewaystable}

\begin{table*}
	 \centering
	\caption{ Comparison between theoretical and experimental \cite{nndc} $\log ft$ values for $^{135}$Te$((7/2^-))$ to the ground and different excited states in $^{135}$I transitions.  Here, symbol `*' indicates the spin parity states proposed by the shell-model  and `1u' denotes 
	first forbidden unique beta decay with log$ft$ = log$f_{1u}t$.}\label{table4} 
%	\begin{ruledtabular}
		\begin{tabular}{lccc|cccccccc}
			\hline
			\multicolumn{4}{c}{~~~~Expt.}&\multicolumn{4}{c}{~~~~~SM}\T\B \\
			\cline{1-4}
			\cline{5-8}
			
			$J_f^{\pi}$ & Energy& $\log ft$& $ {[\overline{C(w_{e})}]}^{1/2}$& $J_f^{\pi}$& Decay mode&$\log ft$ & $ {[\overline{C(w_{e})}]}^{1/2}$      \T\B\\\hline\T\B
			
			$7/2^+$ & 0.0&6.14(2)&26.065 &$7/2^+_1$ & 1st FNU	&6.276 & 22.289\\ \T\B
			$(5/2^+)$ & 603.68(3)&6.45(4)& 18.241&$5/2^+_1$ & 1st FNU	&5.542 & 51.867  \\ \T\B
			$(5/2^+)$ & 870.52(4)&6.55(4)&16.258 &$5/2^+_2$ & 1st FNU	&6.725 &13.284  \\ \T\B
			$(11/2^+)$&1133.21(19) & 9.28$^{1u}$(4)& 0.701 &$11/2^+_1$& 1st FU& $12.093^{1u}$  & 0.028$^{1u}$  \\ \T\B
			$(9/2^+)$	& 1183.86(17)&7.66(5)& 4.530&$9/2^+_1$&1st FNU & 9.058  & 0.906\\ \hline
		\end{tabular}
%	\end{ruledtabular}
\end{table*} 

\end{landscape}

There are also several transitions at a particular energy level where more than one spin parity states are possible experimentally. By analyzing shell-model results of beta decay properties for these transitions, we can confirm one particular spin parity state at that energy level. For instance, in the case of $^{137}$Te$((7/2^-))$ to $^{137}$I$((3/2_1^+,5/2_2^+))$ transition at energy 373.1(7) keV, the experimental $\log ft$ value is 6.6(2) and shell-model $\log ft$ value for $3/2_1^+$ is 11.327 which is very far from the experimental one while this value is 6.435 for spin parity state $5/2_2^+$ which is very close to the experimental one. Thus, the $5/2_2^+$ spin parity state can be suggested at the energy level 373.1(7) keV. 
 It is also mentioned in the Ref. \cite{si} that $5/2^+$ spin parity state can be proposed as the most suitable candidate. 
Similarily, in the case of $^{137}$Te$((7/2^-))$ to $^{137}$I$((3/2_2^+,5/2_3^+))$ transition at energy 600.6(6) keV, the experimental $\log ft$ value is 6.8(1) which is close to the shell-model $\log ft$ value in the case of spin parity state $5/2_3^+$ compared to the spin parity state $3/2_2^+$. Hence, we can suggest a $5/2_3^+$ spin parity state at energy 600.6(6) keV. On moving towards $^{137}$Te$((7/2^-))$ to $^{137}$I$((9/2_2^+,11/2_1^+))$ transition at energy 620.5(10) keV, the shell-model results for spin parity state $9/2_2^+$ are much closer to the experimental $\log ft$ value, i.e., 6.8(1) as compared to spin parity state $11/2_1^+$. Therefore, the spin parity state $9/2_2^+$ can be suggested at energy 620.5(10) keV. Similarly, in the case of $^{137}$Te$((7/2^-))$ to $^{137}$I$((5/2_5^+,9/2_4^+))$ transition at energy 853.2(7) keV, the experimental $\log ft$ value, i.e., 7.0(1) is much closer to the shell-model results for spin parity state $5/2_5^+$ rather than to the spin parity state $9/2_4^+$, therefore, we can suggest $5/2_5^+$ spin parity state at energy 853.2(7) keV. 
Further, at the energy level 955.0(12) keV, the experimental $\log ft$ value is $7.6-9.6^{1u}$ corresponding to $^{137}$Te$((7/2^-))$ to $^{137}$I$((11/2_2^+,13/2_1^+))$ transition whereas this value comes out to be 12.550 for the spin parity state $13/2_1^+$ which is very far from this $\log ft$ value. So, we can suggest the $11/2_2^+$ spin parity state at the energy level 955.0(12) keV. Now, in the case of $^{137}$Te$((7/2^-))$ to $^{137}$I$((5/2_6^+,9/2_6^+))$ transition at energy 1435.9(9) keV, the shell-model $\log ft$ value, i.e., 6.789 for spin parity state $9/2_6^+$ shows quite good agreement with the experimental $\log ft$ value, i.e., 7.1(1). Therefore, the spin parity state $9/2_6^+$ can be suggested at the energy level 1435.9(9) keV. 

Although, the shell-model is showing quite good agreement with the experimental data and is helpful in confirming a particular spin parity state between several unconfirmed spin parity states, but it is failing to confirm a particular spin parity state in some cases as it is giving good results in both of the unconfirmed spin parity states at that particular energy level.

We have also compared theoretical and experimental beta decay properties like $\log ft$ and average shape factor values in the case of $^{135}$Te$((7/2^-))$ to the ground and different excited states in $^{135}$I transitions which are shown in table \ref{table4}. Because there is lack of experimental information on spin parity states at higher energies, we have chosen a limited range of transitions in this work. The value of $g_A^{eff}$ and $\epsilon_{MEC}$ is the same as that of taken in the case of table \ref{table3}. In this case also, the shell-model results are showing good agreement with the experimental data. For example, the experimental $\log ft$ value for the ground state transition from $^{135}$Te$((7/2^-))$ to $^{135}$I$(7/2_1^+)$ is 6.14(2) which is very near to the shell-model $\log ft$ value, i.e., 6.276. Also, in the case of $^{135}$Te$((7/2^-))$ to $^{135}$I$((5/2_2^+))$ transition at energy 870.52(4) keV, the shell-model $\log ft$ value is 6.725, which is close to the experimental $\log ft$ value, i.e., 6.55(4). In some cases, the shell-model $\log ft$ values are not matching well with the experimental ones. This is due to 
unconfirmed spin and parity of these states.
%Thus, the spin parity state $5/2_2^+$ can be the confirmed state at 870.52(4) keV.

\section{Conclusions} \label{secIV}

In this work, we have performed systematic shell-model calculations for nuclear structure properties and beta decay properties of $^{135,137}$Te nuclei and compared the results with the recent experimental data. At first, energy spectra corresponding to $^{137}$I are drawn using jj56pnb effective interaction. On comparing these energy spectra with the experimental spectra, it is inferred that shell-model energy spectra of many spin parity states match well with the experimental data. After that, the dominant wave function configuration is calculated with the help of shell-model calculations. Further, electromagnetic properties corresponding to $^{135,137}$I are being calculated as only magnetic moment value for $^{135}$I$(7/2^+_1)$ is known experimentally, and there is no experimental data available for other electromagnetic observables corresponding to these transitions. Finally, beta decay properties have been calculated corresponding to $^{135,137}$Te (($7/2^-))$ $\rightarrow$  $^{135,137}$I ($J_f$) transitions using the quenched value of weak axial vector coupling constant, i.e., $g_A$=0.66 and mesonic enhancement factor, i.e., $\epsilon_{MEC}=2.0$. It can be concluded that by using the effective value of weak coupling constant and mesonic enhancement factor in $\Delta J=0^-$ transitions, our results for $\log ft$ values improve much better and shell-model calculations for $\log ft$ values show good agreement with the experimental ones.
Further, confirmation of several spin parity states is suggested with the help of shell-model calculations. For instance, $5/2_2^+$, $5/2_3^+$ and $9/2_2^+$ spin parity states are proposed corresponding to experimental energies 373.1(7) keV, 600.6(6) keV and 620.5(10) keV, respectively. Additionally, the spin parity states $5/2_5^+$, $11/2_2^+$ and $9/2_6^+$ are proposed corresponding to experimental energies 853.2(7) keV, 955.0(12) keV,  and 1435.9(9) keV, respectively. Although, the energy spectra with jj56pnb interaction are not in an excellent agreement with the experimental data. On the other hand, the shell-model results for beta decay properties, i.e., $\log ft$ values, are in a good agreement with the experimental data. This implies that the wave functions determined using the shell-model are accurate. Further, the findings of beta decay properties like $\log ft$ values can be enhanced by including core polarization effects. We have also noted that the shell-model calculations for beta decay properties in the $^{132}$Sn region are in order, and it is highly desirable to predict beta decay properties in this region where experimental data is lacking that might be beneficial for future experimental study.

\section*{Acknowledgement}

S. S. would like to thank CSIR-HRDG (India) for the financial support for her Ph.D. thesis work. P. C. Srivastava acknowledges a research grant from SERB (India), CRG/2019/000556. We would like to thanks R. Lozeva, U. K$\ddot{\text{o}}$ster, J. Pore and A. Y. Deo for their valuable suggestions on this work.

\end{document}